\documentclass[12pt,titlepage]{article}

  \renewenvironment{thebibliography}[1]{%
    \begin{oldthebibliography}{#1}%
      \setlength{\parskip}{.35ex}%
      \setlength{\itemsep}{-.04ex}%
  }%
  {%
    \end{oldthebibliography}%
  }

\usepackage{hyperref}
\usepackage{graphicx}
\begin{document}

\begin{center}
{\large \bf CLUSTERS and LARGE-SCALE STRUCTURE:\\ the SYNCHROTRON KEYS}
\vskip 0.1in
Contact author: Lawrence Rudnick, University of Minnesota\\
{\small 116 Church St. SE, Minneapolis, MN 55455. email: larry@astro.umn.edu  }\\
\end{center}
\begin{figure}[h!]
 \begin{center}
\vskip -.07in
\includegraphics[width=4in]{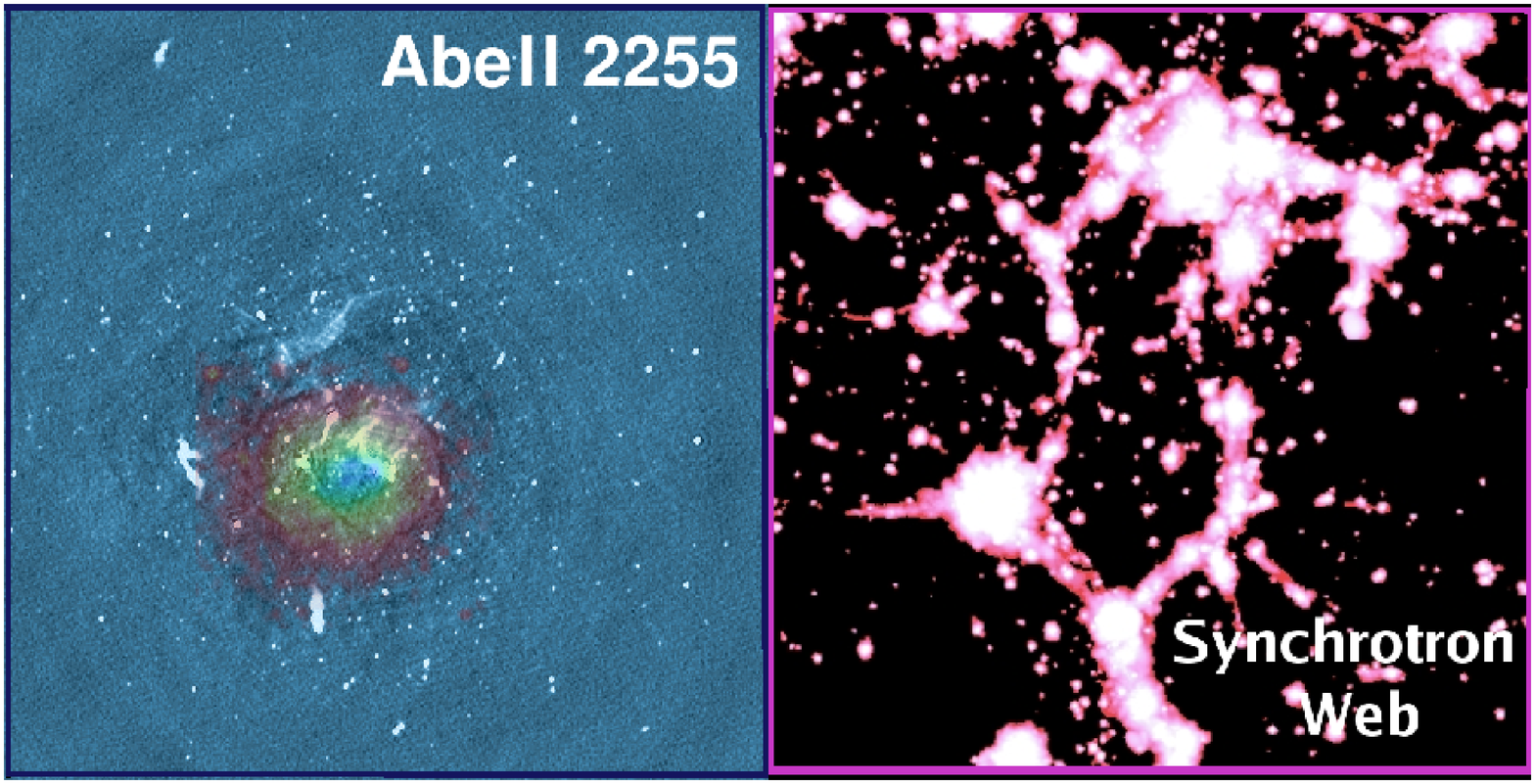}
\vskip 0.01in
{\scriptsize Left: Pizzo et al. 2008, AA 481, 91, (radio) \& I. Sakelliou, MPA (X-ray, colored); Right: Simulation,  K. Dolag, MPA}
\end{center}
\vskip -.2in
\end{figure}
\vskip -.18in
{\small
\vskip -.15in
\begin{tabbing}
Paul Alexander \hskip .75in \= University of Cambridge\\
Heinz Andernach \> Argelander-Institut f\"ur Astronomie \\
Nicholas Battaglia \> University of Toronto \\
Shea Brown \> University of Minnesota \\
Gf. Brunetti \> Istituto di Radioastronomia \\
Jack Burns \> University of Colorado \\
Tracy Clarke \> U.S. Naval Research Laboratory \\
Klaus Dolag \> Max-Planck-Gesellschaft \\
Damon Farnsworth \> University of Minnesota \\
Gabriele Giovannini \> Istituto di Radioastronomia \\
Eric Hallman \> University of Colorado \\
Melanie Johnston-Hollitt \> Victoria University of Wellington \\
Thomas Jones \> University of Minnesota \\
Hysesung Kang \> Pusan National University \\
Namir Kassim \> U.S. Naval Research Laboratory \\
Andrey Kravtsov \> Enrico Fermi Institute, University of Chicago\\
Joseph Lazio \> U.S. Naval Research Laboratory \\
Colin Lonsdale \> Massachusetts Institute of Technology \\
Brian McNamara \> University of Waterloo \\
Steven Myers \> National Radio Astronomy Observatory \\
Frazer Owen \> National Radio Astronomy Observatory \\
Christoph Pfrommer \> Canadian Institute for Theoretical Astrophysics \\
Dongsu Ryu \> Chungnam National University \\
Craig Sarazin \> University of Virginia\\
Ravi Subrahmanyan \> Raman Research Institute \\
Gregory Taylor \> University of New Mexico \\
Russ Taylor \> University of Calgary \\
\end{tabbing}
   }
{\bf \noindent SUMMARY}\\
For over four decades, synchrotron-radiating sources have played a series of pathfinding roles in the study of galaxy clusters and large scale structure. Such sources are
uniquely sensitive to the turbulence and shock structures of large-scale environments, and their cosmic rays and magnetic fields often play important dynamic and thermodynamic roles. They provide essential complements to studies at other wavebands. Over the next decade, they will fill essential gaps in both cluster astrophysics and the cosmological growth of structure in the universe, especially where the signatures of shocks and turbulence, or even the underlying thermal plasma itself, are otherwise undetectable. Simultaneously, synchrotron studies offer a unique tool for exploring the fundamental question of the origins of cosmic magnetic fields. This work will be based on the new generation of m/cm-wave radio telescopes now in construction, as well as major advances in the sophistication of 3-D MHD simulations.


\vskip 0.1in
{\bf \noindent KEY INSIGHTS FROM SYNCHROTRON STUDIES}

\begin{list}{$\bullet$}{
  \setlength{\topsep}{0ex}\setlength{\itemsep}{0ex plus0.2ex}
  \setlength{\parsep}{0.5ex plus0.2ex minus0.1ex}}
\item \textbf{Discovery:~ ~}
      In the early decades, the broad luminosity function of radio galaxies made them the most effective tool for identifying distant clusters  \cite{Smith76}. That discovery mode continues today, e.g., as the high bias of radio galaxies allows identification of 100~Mpc scale superstructures \cite{Brand03}.
\item \textbf{Relativistic/thermal plasma interactions:~~}
    Even before the thermal nature of cluster X-ray emission was confirmed, tailed radio galaxies provided evidence for thermal-like pressures and densities \cite{Miley72,Rudnick76}. Today, AGN outputs create spectacular cavities in the hot plasma and can dramatically influence cluster dynamics \cite{McNamara07}. A separate {\it White Paper} addresses these issues.
\item \textbf{Shocks and turbulence:}
     Dozens of cluster accretion shocks are now visible as synchrotron {\it relics} \cite{Bonafede09} at the peripheries of clusters, providing essential complementary information to model the effects of infall on the thermal gas. Similarly, cluster-wide radio {\it halos} provide evidence for extensive turbulence \cite{Brun08} in the intracluster medium.
\item \textbf{Low density outskirts and filaments:} A handful of radio observations \cite{Brown09,grg} and large-scale simulations \cite{Ryu08} indicate the importance of shocks and turbulence in the low density WHIM, holding out the exciting possibility of using synchrotron emission to illuminate the diffuse component of the cosmic web.



\end{list}
\vskip 0.1in
{\bf \noindent FUTURE SCIENCE CHALLENGES}\\
\vskip -.15in
Progress in understanding cluster astrophysics and the development of large scale structure requires studies across the spectrum, so that we can understand the multiple interactions between dark matter, stars and galaxies, hot thermal gas and the relativistic particles and magnetic fields.  This paper highlight some key areas where synchrotron studies will play a critical role over the next decade.

\indent{\bf $\bullet$ ~What is the dynamical state of the baryons in clusters and lower density WHIM filaments as they evolve over cosmological timescales?}\\
\indent{\it ~ ~ ~ What are the regulating roles of {\bf shocks, turbulence and the relativistic plasma}}?


\indent {\bf $\bullet$~What is the cosmological origin of magnetic fields?}\\

\vskip -.15in
\noindent{\bf A. Dynamical State of Cluster Baryons ~ }  Synchrotron studies provide unique diagnostics of shock structures as well as turbulence on all scales in the ICM.  When these structures are weak, e.g., from minor or previous activity, synchrotron signatures become essential. In addition, the relativistic plasmas play an active role in the evolving thermal plasma, seen most directly through AGN outflows, but also throughout the cluster medium. Magnetic fields can also have substantial influences on the thermal properties of the ICM \cite{Maron03} and stabilize cavities against  Kelvin-Helmholtz and Raleigh-Taylor instabilities \cite{Dursi08}. 
\begin{figure}[ht]
\begin{center}
  \includegraphics[width=5in, height=2.55in]{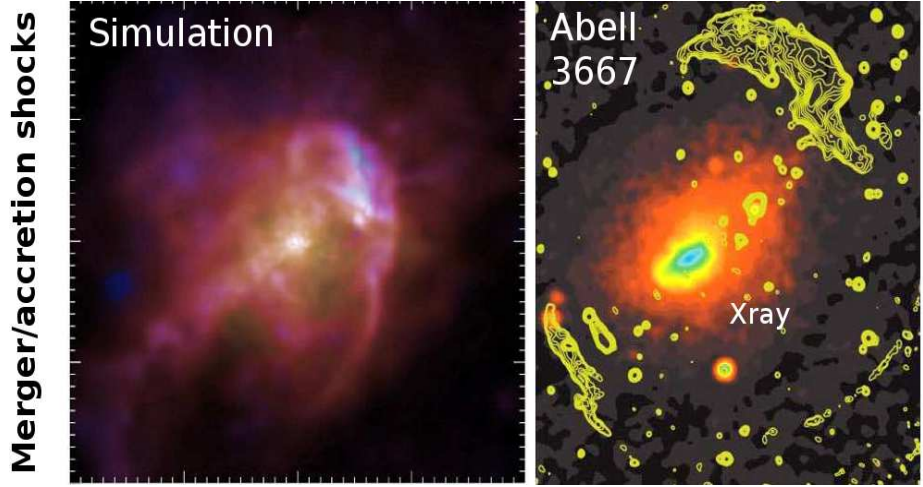}
\end{center}
\vskip -.15in
  \caption{\label{shocks} {\sc synchrotron illumination of shocks.}~ ~ {\small Left: Mach number distribution from cluster simulation \cite{Battaglia08}; Right: Contours of synchrotron accretion shocks \cite{Melanie}.}}
\end{figure}
\vskip -.08in
Figure \ref{shocks} shows the promise of synchrotron studies for tracing the formation and continued evolution of clusters through mergers and accretion.  The next generation of radio telescopes will provide exquisite images of multiple weak shock structures in individual clusters \cite{Battaglia08}, allowing us to characterize their accretion history and the eventual dissipation of the shocks and turbulence.  Simulations have become increasingly successful in reproducing the observed radio properties through models of turbulent field amplification and cosmic ray acceleration at shocks, giving confidence that shocks and turbulence essentially invisible by other means may be effectively measured \cite{Pfrommer07,Markevitch07,Skillman08}. The thermal and non-thermal components are strongly correlated. Maps of the radio spectral index distribution \cite{A2163}, e.g.,  have shown that regions with more energetic relativistic particles are coincident with regions with high gas temperature. Dramatic advances in shock physics and cosmic ray generation and confinement are also expected with new {Fermi} and {Cerenkhov} measurements.

\begin{figure}[h]
\begin{center}
  \includegraphics[width=6.5in, height=2.2in]{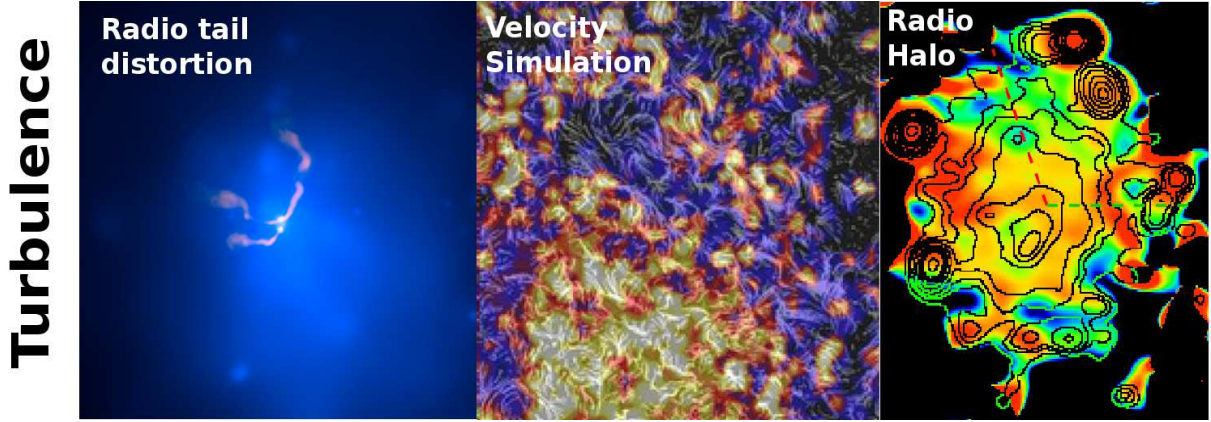}
\end{center}
  \caption{\label{turbulence} {\sc probing cluster turbulence.}~ ~{\small 
    Left:  Abell 400 with tailed radio galaxy 3C75, buffeted by ICM motions \cite{3C75}. Middle: simulation results on cluster velocity turbulence \cite{Dolag05}. Right: Spectral index variations in Abell~2163 cluster halo (N to left) \cite{A2163}}.}
\vskip -0.15in
\end{figure}

There is an increasing recognition of the important  role of turbulence in the hot thermal plasma \cite{Schuecker04}. Figure \ref{turbulence} illustrates how synchrotron observations provide signatures of turbulence, through the distortion of flows in tailed radio galaxies and in the production of the very steep radio spectra and inhomogeneities of cluster-wide radio halos.  Turbulence changes the  distribution of entropy, and in cool-core clusters, the distribution of metals in the ICM \cite{Takizawa05,Scannapieco08}. The diffuse relativistic plasma can directly contribute significant energy and pressure to the ICM, which will substantially affect its evolution.


\vskip .08in
\noindent{\bf B. Structure and dynamics of the WHIM ~  }

\begin{figure}[h]
\begin{center}
  \includegraphics[width=6.5in, height=1.9in]{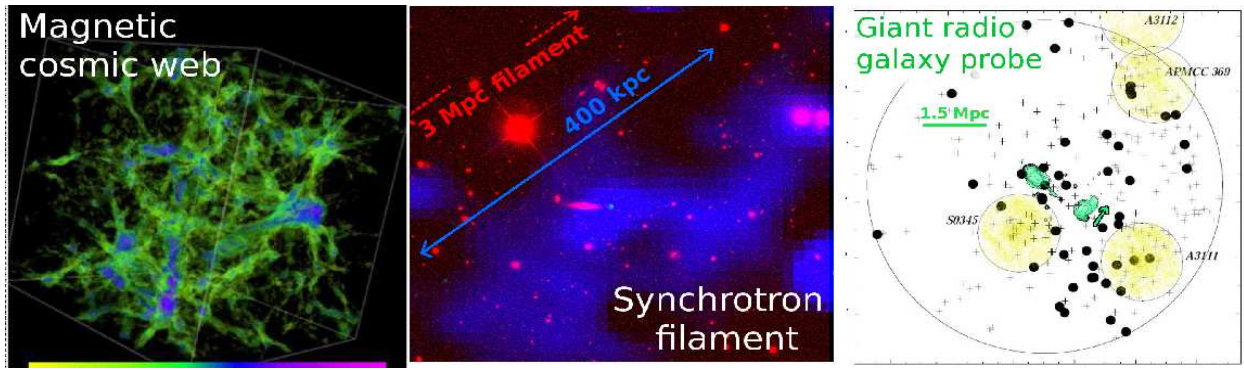}
\end{center}
 \vskip -.15in
 \caption{\label{lowdens} {\small {\sc probing large scale structure.}
     100~Mpc box of magnetic field structures \cite{Ryu08}. Middle: Synchrotron detection of possible cosmic filament \cite{Brown09}. Right: Distortion (green arrow) of Giant Radio Galaxy  by (invisible) diffuse medium between (yellow) clusters \cite{grg}}
     }
\vskip -.1in
\end{figure}

Nonthermal probes of the WHIM are a critical complement to the extremely difficult thermal
emission/absorption line studies \cite{Tripp08}. In the future, Generation-X X-ray satellites may reach the sensitivities needed for diffuse thermal emission. Today, however, synchrotron emission from large-scale filaments is on the threshold of detection.  Shocks from infall onto and along filamentary structures accelerate cosmic rays, and the post-shock turbulent flows can be extremely effective in amplifying magnetic fields, while producing pressures exceeding the thermal contributions \cite{Ryu08}.  A variety of simulations therefore hold out the exciting possibility for detecting the diffuse WHIM in synchrotron radiation, \cite{Keshet04,Dolag06}, although the emissivity levels are still dependent on the unknown origins of seed magnetic fields and the pre-injection of cosmic rays by past AGNs. As these issues are sorted out, synchrotron-based estimates of WHIM density and pressure will become possible. A handful of observations indicate that emission from intercluster filaments has already been seen \cite{Brown09}; dramatically new and improved telescope facilities may allow us to illuminate the ``cosmic web'' in the coming decade. Another powerful probe of the diffuse components outside clusters comes from distortions seen in Giant ($>$1~Mpc) Radio Galaxies (Figure \ref{lowdens}, \cite{grg}).

Synchrotron studies will also make important contributions to structure development tests of {\bf $\Lambda$CDM} cosmogony.  Steep spectrum radio emission may be used to identify cosmologically important cases such as the Bullet Cluster, and, for dark energy studies, separate relaxed systems from recent mergers \cite{Allen04}, (which are expected to increase greatly at higher redshifts) {\it even when no X-rays are visible}. They will also trace the evolution of the diffuse baryons in large scale structure.


\vskip .075in
\noindent{\bf C. Magnetic field origins ~ ~ } \\
The cosmic origin of large scale magnetic fields \cite{Zweibel06}
 is a long-standing, fundamental problem.  These fields, and their coupling to the cosmic rays are
 important since they control particle diffusion and transport (viscosity, conduction)  \cite
 {Schek05}, and deflect ultra-high
 energy cosmic rays, thus limiting our ability to retrace UHECR
 arrivals back to their origins \cite{Abraham08}. In clusters, the extensive amplification and modification of field geometry erases the signatures of their origins. Magnetic field studies of the WHIM, through polarized emission and Faraday rotation of background sources therefore holds the best promise for addressing this key issue.
\vskip .1in
\noindent {\bf EMERGING TOOLS}\\
\vskip -.15in
{\bf Telescopes:~ ~}   Major new cm-m wavelength telescope facilities, including the EVLA, the new multi-beam systems at Arecibo, the Long Wavelength Array, the GMRT, LOFAR, ASKAP, MeerKAT, MWA and eventually the SKA promise to revolutionize synchrotron studies of clusters and large scale structure.  The current generation of radio telescopes has provided images of the diffuse radio
emission from only the brightest $\sim$50 clusters of galaxies \cite{Ferrari08}. \emph {Based on the current luminosity functions alone}, the new low frequency arrays will be sensitive to at least 10$^3$ halos \cite{Brun08}, especially by capitalizing on the steep spectra. This will allow, for the first time, the separation of contributions from the various cluster properties including X-ray mass and temperature structure, accretion/merger history, cosmic epoch, etc.  For the first time, multiple internal shocks can probe past mergers and AGN outbursts. Together with X-ray measurements, we will begin to reconstruct the thermal history of the ICM.  Vast new populations of accretion shocks and lower mass cluster sources will be detected, especially in environments currently undetectable to X-rays.  These new radio surveys, and hopefully next generation X-ray facilities will allow us to begin mapping the diffuse ``cosmic web''.

{\bf Simulation Technology:~ ~} Rapidly increasing capabilities in numerical simulations have allowed much more physically realistic hydrodynamic and magneto-hydrodynamic modeling of clusters and large-scale structure.  Simulations are increasingly better tied to observational predictions, e.g., for emissivity, polarization and Faraday rotation, X-ray emission, cosmic ray production, etc. \cite{Battaglia08}. In the next generation, higher resolution studies will allow the accurate tracking of shocks and turbulence in clusters and filaments, while being anchored in cosmologically relevant structures, e.g., the ``Millennium Simulation''.
\vskip .075in
\noindent{\bf EXAMPLE KEY PROJECTS}\\
\vskip -.05in
\noindent {\bf RADIO SKY SURVEYS for DIFFUSE CLUSTER EMISSION:~ ~ }
 The goal of these surveys is to increase by 1-2 orders of magnitude the number of radio halos, ``peripheral relic'' accretion shocks, as well as to produce the first extensive evidence for shocks associated with the low density WHIM. Major advances will be made by LOFAR and other SKA-prototypes.  At redshifts $\le$0.2, a combination of sensitive single dish (GBT, Arecibo) and EVLA measurements, e.g., at 21cm will allow a clean separation of background confusion, weak shocks, and more diffuse turbulent generated emission not only in clusters, but in their outskirts. Addition of a compact E-configuration to the EVLA would dramatically increase its utility for diffuse emission studies. Survey projects already being planned for the upcoming EVLA will push these studies to higher redshifts, utilizing an all sky, full Stokes, low-resolution survey akin to the NVSS 
(rms sensitivity of 22~$\mu$Jy/beam, 45'' resolution)

Because of the steep radio spectrum of diffuse cluster-related sources, lower frequency instruments such as the U.S. LWA are critical for opening up the discovery space in the next decade \cite{Cassano08}. The LWA covers the range of 20 to 80 MHz  with a large field of view ($2\times (80MHz/\nu$) degrees), excellent surface brightness
sensitivity ($\Sigma \sim 1$ mJy/bm) and high angular
resolution ($2\times (80MHz/\nu)$ arcseconds).  Targeted and blind surveys and followup spectral studies of steep spectrum cluster halos and relics with the LWA will be most valuable, providing information on shock strengths and structures, merger diagnostics, and  diffusive shock and other relativistic particle acceleration processes. The LWA will also be a discovery tool for the emerging class of  smaller radio ``mini-halos''
associated with cooling core clusters. Looking toward the end of the next decade and beyond,
  the SKA will ultimately provide both much higher sensitivity and a larger field
  of view.

\vskip 0.2in 
\noindent {\bf {CLUSTER MAGNETIC FIELD TOMOGRAPHY:}}
Rotation measures of both background sources seen through clusters and RM variations across sources embedded in clusters can be used to measure key parameters of the turbulent power spectrum in cluster fields, including the injection and coherence scales and type of turbulence.    At present, dedicated small area surveys can produce a extragalactic background count of $>$40/deg$^{2}$ \cite{Taylor07} suitable for statistically constraining
magnetic fields in super-cluster regions \cite{Xu06}. A deeper 
EVLA \emph{cluster} survey will provide much finer sampling of extragalactic background RMs 
($<$200$^{\prime\prime}$ separation), enabling the study of  individual 
clusters out to redshifts $<$0.5, and sufficient to test the turbulent MHD 
magnetic field amplification models of WHIM filament emission \cite{Ryu08}. 
\begin{figure}[ht]
\begin{center}
  \includegraphics[width=4in, height=2.5in]{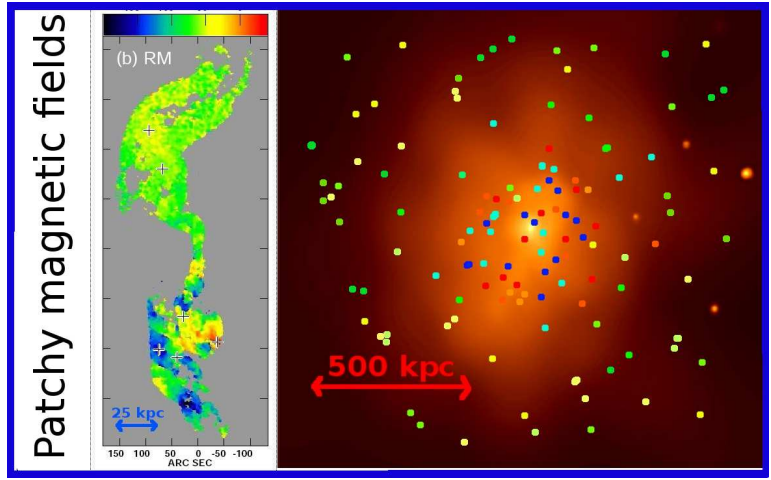}
\end{center}
\vskip -.09in
  \caption{\label{sija} 
     {\sc mapping cluster turbulence.}~ ~Left: Four frequency rotation measure determinations across radio galaxy 3C31 in NGC383 group from \cite{Laing08}. Right:  Cartoon illustrating increased scatter in RMs of background sources behind rich cluster. (X-ray image of richness class 4 cluster from \cite{Arab02}).}
\end{figure}
High resolution RM maps of cluster shocks (or embedded extended radio 
galaxies) have the power to reveal the nature of MHD turbulence, e.g., the 
power spectrum can discriminate between Kolmogorov ($k^{-5/3}$) and 
Burgers ($k^{-2}$) turbulence spectra \cite{Battaglia08}.  We must 
sample $\sim$10 kpc scales for this measurement (or 5'' at z=0.1), which can be reached by an
EVLA deep cluster survey.

\vskip 0.2in

\noindent {\bf{SYNTHETIC CLUSTERS and LARGE SCALE STRUCTURE:}} 
The ultimate goal of the next generation of computer simulations is to make them true to our best understanding of all relevant physical processes and almost indistinguishable from observations in all wavebands. Our capability to model diffuse radio emission from galaxy clusters
has increased immensely through, e.g., coupling cosmological hydrodynamic simulations
with semi-analytical modeling of magnetic fields and
relativistic electron distributions \cite{Pfrommer08}. Particle-in-cell simulations are leading the way in
studying particle acceleration via the Fermi process but are not yet
scaled to cluster sizes or timescales. These simulations will allow one to
better calibrate and advance the current semi-analytical models that are
implemented into cosmological simulations. In order to completely model
the intracluster medium there are additional sources of relativistic
particles such as active galactic nuclei and turbulence as well as the
reservoir of mildly relativistic particle that are necessary to include.
Also, further improving the cosmological simulations so that one can
resolve individual galaxies and shock fronts within a “mini universe” and
the inclusion of magneto-hydrodynamics will ensure that one has included the
essential physical processes and scales to sufficiently study diffuse
radio emission.

These enormous undertakings will demand that  "Top-100" computational facilities
be available to many research groups and infrastructure designed so that
results are easily accessible to theorists and observers for comparison
and analyses.
\vskip 0.125in
\noindent {\bf CONCLUDING POINTS}
\vskip 0.075in
$\bullet$ The diffuse baryons in clusters and in the lower-temperature filaments are in a highly dynamical state, driving the continuing evolution of both large scale structure and galaxies. \
\vskip .05in
$\bullet$ Synchrotron studies play unique roles in determining the influence of shocks and turbulence in the thermal medium, and quantifying the direct physical effects of the relativistic plasma.  In the cooler WHIM regions, synchrotron studies appear very promising for illuminating the diffuse cosmic web.
\vskip .05in
$\bullet$ Observations across the electromagnetic spectrum are critical to address the evolution of diffuse baryons.  Existing and new UV (COS), X-ray (IXO, and eventually Gen-X) and gamma-ray (Fermi and Cerenkov) facilities, along with the new generation of S-Z instruments (South Pole Telescope, SZA) are essential to the science, and must be supported.  Support for the EVLA and the new generation of low frequency, SKA-precursor arrays is likewise essential for characterizing the diffuse medium and its evolution. Comprehensive simulations with realistic physics and ``synthetic'' observations in all wavebands are needed to understand what we see.

\end{document}